\documentclass[aps,prx,showpacs,amsmath,amssymb,amsfonts,superscriptaddress,onecolumn,longbibliography]{revtex4-1}
\usepackage{graphicx}
\usepackage{verbatim}
\usepackage{dcolumn}
\usepackage{bm}
\usepackage{color}
\usepackage[colorlinks=true,citecolor=blue,linkcolor=blue]{hyperref}

\begin{document}

\title{Modeling bacterial flagellar motor with new structure information: Rotational dynamics of two interacting protein nano-rings} 

\author{Yuansheng Cao}
\affiliation{Department of Physics, University of California, San Diego, La Jolla, CA, United States }

\author{Tairan Li}
\affiliation{Yuanpei College \& Center for Quantitative Biology, Peking University, Beijing, China}

\author{Yuhai Tu}
\affiliation{Yuhai Tu, IBM T.J Watson Research Center, Yorktown Heights, New York 10598, USA}
\email{yuhai@us.ibm.com}



\begin{abstract}

In this article, 
we develop a mathematical model for the rotary bacterial flagellar motor (BFM) based on the recently discovered structure of the stator complex (MotA$_5$MotB$_2$). The structure suggested that the stator also rotates. The BFM is modeled as two rotating nano-rings that interact with each other. Specifically, translocation of protons through the stator complex drives rotation of the MotA pentamer ring, which in turn drives  rotation of the FliG ring in the rotor via interactions between the MotA ring of the stator and the FliG ring of the rotor. Preliminary results from the structure-informed model are consistent with the observed torque-speed relation. More importantly, the model predicts distinctive rotor and stator dynamics and their load dependence, which may be tested by future experiments. Possible approaches to verify and improve the model to further understanding of the molecular mechanism for torque generation in BFM are also discussed.  
\end{abstract}

\maketitle
\newpage

\section{Introduction}

The swimming motion of many bacteria is powered by the rotary bacterial flagellar motor (BFM). As a canonical example, the motors in {\it E. coli} are fueled by eletrochemical potential difference of protons ($H^+$) across the cytoplasmic membrane, i.e., the proton motive force (PMF)[\cite{ Larsen1974,Hirota1981,HBerg2003}]. The BFM can rotate bidirectionally. Without binding to the phosphorylated response regulator CheY protein (CheY-P), the BFM rotates in a counter-clockwise (CCW, viewed from the filament to the motor) direction. There are $\sim 4-6$ BFMs in an {\it E. coli} cell. The CCW rotation of the BFMs causes their flagellar filaments to form a coherent bundle and propels the swimming cells in a roughly straight line that is called a ``run." However, binding of CheY-P to BFM causes it to rotate in the clockwise (CW) direction, which disrupts the filament bundle. As a result, the rod-shaped {\it E. coli} cell changes its orientation randomly without translational movement in an event called a ``tumble".

The BFM consists of a rotor and a number of stators that can vary from $0$ to $11$ depending on gene expression and mechanical load [\cite{blair1988,Lele2013dynamics,tipping:2013:mbio,nord:2017:catchbond,wadhwa:2019:pnas,wadhwa:2021:pnas}]. The rotor consists of FliF, FliG, FliM and FliN. FliF forms the MS-ring embedded within the cytoplasmic membrane [\cite{Minamino2014,Lee2016,Asai1997,Sato2000,Yorimitsu2004}]. The cytoplasmic face of the MS-ring is attached to the C-ring , which comprises the FliG, FliM and FliN proteins. The diameter of the C-ring is $\sim 45nm$. The C-ring is responsible for coupling the rotor to the stator units and for switching of rotational directions. The membrane-bound stators, which consist of the MotA and MotB proteins, are powered by the PMF to drive the rotation of the rotor via direct interaction between MotA and FliG [\cite{Blair1990,kojima2004s,Chun1988,Roujeinikova2008}]. 

The molecular mechanism of torque generation by the stator units has been suggested by two recent Cryo-EM studies of the stator structure [\cite{deme2020,santiveri2020}]. Both studies report that a stator unit consists of a MotB dimer surrounded by a MotA pentamer ring with a diameter $\sim 5-7.5nm$. The position of the MotB dimer, which is bound to the peptidoglycan layer in the rigid cell wall, remains fixed, while the MotA pentamer forms a ring that can rotate around the MotB dimer. Each MotB monomer progressively engages with each MotA monomer to form an ion channel through which one proton can pass to drive the rotation of the MotA pentamer ring. 

Brownian ratchet models have attracted physicists’ attention for modeling molecular motors since Richard Feynman popularized the idea more than half a century ago [\cite{Feynman1966,Parrondo1998,Astumian2010,Peliti2012}]. Among all variants of the ratchet models, only isothermal chemical ratchets are relevant for biological motors [\cite{Julicher1997,Parrondo2002}]. A minimal model for rotational ratchet model can be found in [\cite{Xing2006,Meacci2009,Meacci2011}]. This model has two types of coordinates: the ``physical" coordinate that describes the motion of the motor in physical space (angles for rotary motor); and the ``chemical" coordinate that describes the different conformational states of the motor.
For the BFM, the motor moves in a periodic potential in which torque, which is the gradient of the potential, is generated as the motor slides down from a position of high potential energy towards its equilibrium position, which has the lowest potential energy. The passage of an ion triggers a conformational change (stepping) of the motor. Because the landscape of potential energy is different for different conformational states of the motor, the conformational change driven by the PMF changes the potential energy landscape and can bring the motor to a position of high potential energy in the new potential energy landscape. The newly gained potential energy continues to drive the physical rotation of the motor. This continuous process drives the system towards a sequence of new equilibrium positions and gives rise to a directed stepwise rotation [\cite{Sowa2005}].

One of the most important characteristics of a rotary motor is its torque-speed dependence (or force-speed dependence for linear motors). The measured torque-speed curve for the CCW BFM has a downward concave shape with a roughly constant high torque at low to medium speeds and a rapid and nearly linear decrease of torque at high speeds [\cite{Chen2000,Lo2013}]. This concave torque-speed curve has the advantage of generating high power output (or, equivalently, a high torque for a given speed) in a wide range of physiologically relevant loads. The minimal ratchet model can reproduce the observed torque-speed curve when certain general constraints for the rotor-stator interaction potential and the proton-assisted transition rate between different conformational states are applied [\cite{tu2018design}].

Before the recent availability of an accurate picture of the structure of the stator [\cite{deme2020,santiveri2020}], previous modeling studies assumed that the stator unit has a MotA$_4$MotB$_2$ composition[\cite{kojima2004s,nirody2017}]. 
The newly discovered structure of the stator complex has the different stoichiometry of MotA$_5$MotB$_2$ [\cite{deme2020,santiveri2020}]. The structure of the stator complex immediately led to an important new insight about the molecular mechanism by which the BFM operates: the MotA ring in the stator unit rotates.  This new insight means that the stator unit must be described by its physical state in terms of the relative angle between the MotA pentamer ring and the MotB dimer in addition to its chemical (conformational) states. In the following section, we present an updated model for the BFM based on the general modeling framework described before [\cite{tu2018design}], but now in the context of the new structure of the stator. Some preliminary results predicted by the model will also be described.

\section{A structure-informed model and some preliminary results}
\begin{figure}[h!]
\begin{center}
\includegraphics[width=17cm]{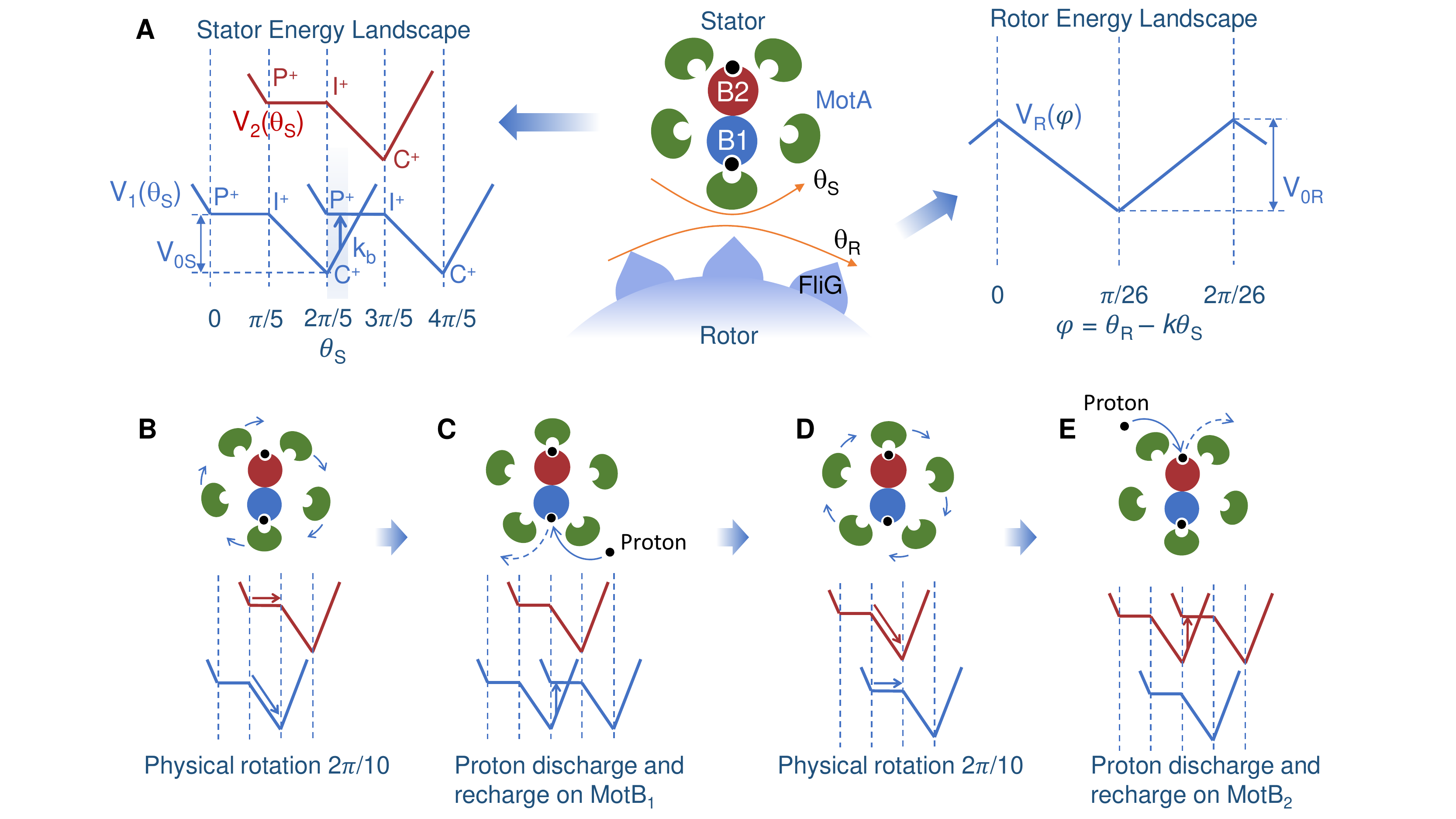}
\end{center}
\caption{ A structure-informed model for the BFM. (A) Illustration of the two-rotating-ring model for the BFM (center panel). 
The stator energy landscape (left panel) shows the two interaction potentials between the MotA pentamer ring and the MotB dimer: $V_1(\theta_S)$ from MotB$_1$ (blue) and $V_2(\theta_S)$ from MotB$_2$ (red), which have the same shape but are shifted from each other by half a period $\Delta\theta_S/2=2\pi/10$. 
$P$ and $C$ stand for the two conformational states of MotB in which the proton acceptor residue faces the periplasm and the cytoplasm, respectively, and $I$ stands for an intermediate state between P-state and C-state. The plus (or minus) superscript indicates the proton-bound (or proton-unbound) state. 
Transition (stepping) from one MotA-MotB interaction potential to another interaction potential occurs with a probability rate $k_b$ near the potential minimum (shaded region in the left panel), which is driven by the proton discharge-recharge chain reaction $P^+\to P^-\to C^-\to C^+$. For brevity, only the initial and final states ($P^+$ and $C^+$) are shown, and the intermediate discharged states $C^-$ and $P^-$ are omitted (see text for the full description). The rotor energy landscape (right panel) shows the interaction potential ($V_R$) between MotA and FliG, which depends only on the scaled angle difference $\varphi=\theta_R-k\theta_S$. For simplicity, we used a piece-wise linear form for $V_{1,2}$ and $V_R$. To the left of the $P\to I$ plateau in $V_{1,2}$, a barrier is added to prevent back flow. The four stages of stator dynamics during one full power stroke as described in Eq.~\ref{chain_reaction} are illustrated in B-E.
(B) Physical rotation of the MotA pentamer ring, which is driven dominantly by MotB$_1$. (C) Proton (the black dots) discharge and recharge on MotB$_1$, which causes the transition of the MotA-MotB$_1$ interaction potential and effectively shifts $V_1$ (blue line) by one period ($\Delta\theta_S$). (D) Physical rotation of the MotA pentamer ring, which is driven dominantly by MotB$_2$. (E) Proton discharge and recharge on MotB$_2$, which shifts the MotA-MotB$_2$ potential $V_2$ (red line) by one period. 
}
\label{fig:1}
\end{figure}
Our new model describes the dynamics of the two coupled rotating nano-rings (the MotA ring and the FliG ring) as shown in Fig.~1A. The stator unit is the active part of the motor. It can generate torque as a rotational ratchet powered by the PMF. Rotation of the MotA ring drives the rotation of the passive FliG ring through a coupling potential $V_R(\theta_S,\theta_R)$, where $\theta_S$ and $\theta_R$ are the rotational angles of the stator and the rotor, which have periodicity of $\Delta\theta_S=2\pi/5,\Delta\theta_R=2\pi/26$, respectively. (It is assumed that there are $26$ FliG proteins in the C-ring [\cite{Thomas2006}].) The rotational motion of the rotor is described by a Langevin equation for the rotor angle $\theta_R$: 
\begin{equation}
    \xi_R\frac{d\theta_R}{dt}=-\frac{\partial V_R(\theta_S,\theta_R)}{\partial\theta_R}+\eta_R(t),
    \label{rotor}
\end{equation}
where $\xi_R$ is the load on the rotor, the first term on the right hand side of Eq.~\ref{rotor} is the torque exerted on the rotor from the stator via the MotA-FliG interaction, and $\eta_R(t)$ is the rotor thermal noise: $<\eta_R(t)\eta_R(t^{\prime})>=2k_BT\xi_R\delta(t-t^{\prime})$ with $k_B$ the Boltzmann constant and $T$ the temperature. The coupling potential $V_R(\theta_S,\theta_R)$ should be periodic in both $\theta_S$ and $\theta_R$. To satisfy this dual periodicity, we write $V_R(\theta_S,\theta_R)=V_R(\varphi)$ where $\varphi=\theta_R -k \theta_S$ is the scaled relative angle, with $k=\Delta\theta_R/\Delta\theta_S=\frac{5}{26}$, which is the gear ratio between the two rotating rings. 
 
We model the stator unit as a rotational ratchet driven by the PMF. Based on the new structural information, the MotA pentamer ring rotates around the fixed MotB dimer. Given the incommensurate stoichiometry ($5:2$) within the stator unit, the two MotB monomers engage with MotA monomers alternately -- when one MotB is engaged directly with one MotA monomer the other MotB is positioned between two MotA monomers. Consequently, the two MotB monomers generate two different energy potentials shifted by half a period ($\Delta \theta_S/2$). To account for this effect, the overall stator potential can be written as the sum of two identical periodic potentials shifted by an angle difference of $\Delta\theta_S/2$: $V_S(\theta_S)=V_1(\theta_S)+V_2(\theta_S)$ with $V_2(\theta_S)=V_1(\theta_S+\Delta\theta_S/2)$ as shown in Fig.~1A.

During one power stroke, the MotA pentamer ring rotates one period ($2\pi/5$). Each full power stroke consists of two half strokes, each of which is powered by a proton passing through one of the two ion channels alternately. During each half stroke, the proton acceptor residue (D32 for {\it E. coli}) on the corresponding MotB monomer becomes protonated, which drives rotation of the MotA ring. At the same time, the protonated MotB undergoes a conformational change in which the proton acceptor residue changes from a periplasm-facing position to a  cytoplasm-facing position, from which the proton can be released to the cytoplasm. To describe this process, we define three conformational states for each MotB: a periplasm-facing state called the P-state, a cytoplasm-facing state called the C-state, and an intermediate state between the P-state and C-state, which we call the I-state. Each MotB can be either protonated or unprotonated (indicated by a plus or minus sign). A proton can bind with MotB when it is in the P-state; and a bound proton can enter the cytoplasm when the MotB is in the C-state.  

As suggested by the recent Cryo-EM study [\cite{santiveri2020}], different conformational states occur at different relative positions between MotB and the closest MotA monomer, or, equivalently, at different locations within the MotA-MotB landscape of interaction energy potential, as shown in Fig.~1A (left). In particular, the P-state occurs near the peak potential and the C-state occurs near the minimum. For simplicity, we used a simple piece-wise linear potential ($V_{1,2}$) that is flat over half of the period and linear in the other half, as shown in Fig.~1A (left). Using this simple representation of the internal chemical states of the stator, evolution of the internal states of the two MotB monomers in a stator unit during one full rotational period ($2\pi/5$) can be described by the reaction cycle:   
\begin{equation}
P^+_2\underbrace{I_1^+\to I^+_2}_{\text{Rotation}}\underbrace{C_1^+ \to I_2^+C_1^- \to I_2^+P_1^- \to I_2^+}_{\text{Proton discharge and recharge on MotB$_1$}} \underbrace{P_1^+\to C^+_2}_{\text{Rotation}}  \underbrace{I_1^+\to C_2^- I_1^+ \to P_2^- I_1^+ \to P_2^+ }_{\text{Proton discharge and recharge \\ on MotB$_2$}}I_1^+, \label{chain_reaction}
\end{equation}
where the subscripts (1,2) indicate the two MotB monomers. For each MotB, torque is generated during the transitions $P^+ \to I^+ \to C^+$, which drives the physical rotation of the MotA ring. After reaching the $C^+$-state near the potential energy minimum, MotB is rejuvenated by the release of the proton to the cytoplasm followed by a conformational change from the C-state to the P-state and a subsequent addition of a new proton to the P-state: $C^+ \to C^- \to P^- \to P^+$. This chain reaction returns MotB to its high-energy protonated P-state ($P^+$) to continue driving the rotation of the stator. It is clear from the reaction cycle (Eq.~\ref{chain_reaction}) that one full power stroke is powered by translocation of two protons (or 2PMF), each ``consumed" by one MotB. The four stages of the alternating physical rotations and proton discharge-charge reactions during one full power stroke as described in Eq.~\ref{chain_reaction} are illustrated in Fig.~1B-E, respectively.

In our model, the rotation dynamics of the MotA ring can also be described by a Langevin equation for the stator angle $\theta_S$:
\begin{equation}
    \xi_S\frac{d\theta_S}{dt}=-\frac{\partial V_R(\theta_S,\theta_R)}{\partial\theta_S}-\frac{\partial V_S(\theta_S)}{\partial\theta_S}+\eta_S(t),
    \label{stator}
\end{equation}
where $\xi_S$ is the load (viscosity) of the stator, the first and the second term on the right hand side of Eq.~\ref{stator} are the torque exerted on the stator from the rotor via the MotA-FliG interaction and the torque generated by the MotA-MotB interaction, respectively, and $\eta_S(t)$ is the stator thermal noise with $<\eta_S(t)\eta_S(t^{\prime})>=2k_BT\xi_S\delta(t-t^{\prime})$. The shape of $V_S(\theta_S)$ is given in Fig.~1A. To prevent back flow at the potential energy maximum, we introduce an additional energy barrier which extends to the left of the $P\to I$ plateau. 

The key chemical reaction that powers rotation of the stator is the proton discharge-recharge chain reaction ($C^+ \to C^- \to P^- \to P^+$), which leads to the transition (stepping) from one potential landscape, in which the system has a low (near minimum) potential energy, to another potential landscape where the system has a high (near maximum) potential energy. For simplicity, this chain reaction is modeled here by an overall transition rate $k_b(\theta_S)$ that depends on the MotA-MotB relative angle $\theta_S$. (See Fig.~1A for an illustration of $k_b$, the blue arrow in the left panel.) Specifically,
\begin{equation}
    \text{Prob(transition from low to high potential energy during $t\to t+dt$)}=k_b(\theta_S)dt. 
    \label{chem}
\end{equation}

In summary, our model considers the coupled rotations of both the active stator and the passive rotor as well as the key chemical reactions that power the rotation of the stator. The dynamics of the BFM are fully described by Eqs.\ref{rotor},\ref{stator} and \ref{chem}, which depend on three biologically meaningful functions: two interaction energy (potential) functions - $V_R(\theta_S-k\theta_R)$ between the FliG-ring of the rotor and the MotA ring, $V_S(\theta_S)$ between the MotB dimer and the MotA ring, and the chemical transition rate function $k_b(\theta_S)$. Next, we present some preliminary results based on simulations of the model with simple choices for these functions to demonstrate the general utility of the model. 


\begin{figure}[h!]
\begin{center}
\includegraphics[width=17cm]{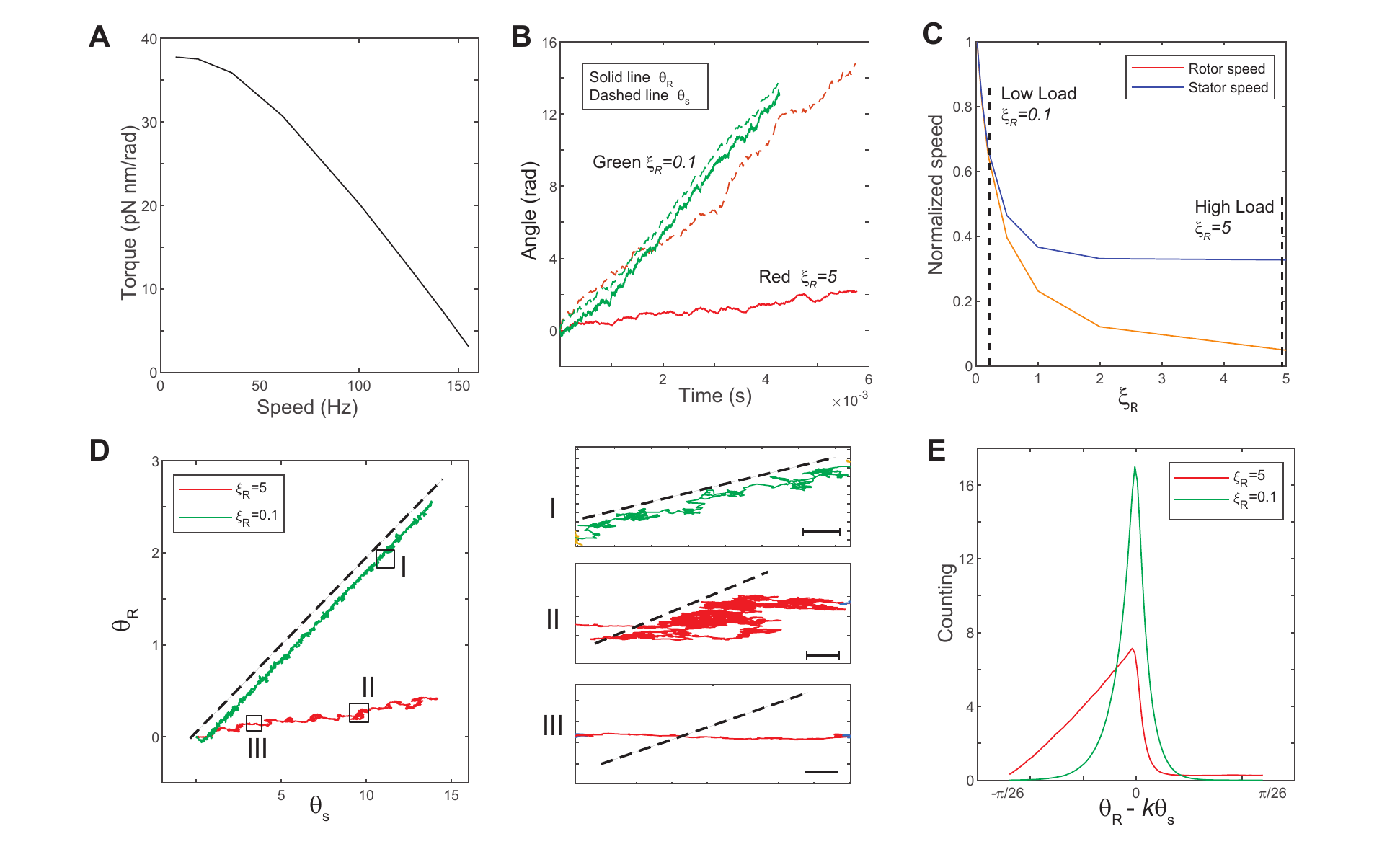}
\end{center}
\caption{Model results. (A) The torque-speed curve from the model with simple choices of the interaction potentials ($V_R$ and $V_S$) and the transition (stepping) rate $k_b$ (parameters are given at the end of the caption).  (B) Time series of $\theta_R$ (solid line) and $\theta_S$ (dashed line) for different rotor loads. Red: high load $\xi_R=5$, green: low load $\xi_R=0.1$. (C) The normalized average speeds of the rotor (yellow) and stator (blue) versus the rotor load $\xi_R$. (D) Left: Trajectories in $(\theta_R,\theta_S)$ space for different loads. Dashed black line: $\theta_R=k\theta_S$, with $k=\Delta\theta_R/\Delta\theta_S=5/26$. Right: three typical movements with zoom-in trajectories. I: rotor moves along with stator without slipping; II: reduced net speed for both rotor and stator; III: stator rotates while rotor speed is near zero. The black dashed lines indicate the no-slip motion $\theta_R=k\theta_S$. Scale bar: 0.3 rad.  (E) The distributions (histogram) of $\varphi=\theta_R-k\theta_S$ for a high load and a low load. Piece-wise linear forms for $V_R$ and $V_{1,2}$ are used. $V_{1,2}$: $k_BT=4.11 pN\times nm$, $V_{0S}=10k_BT$, $\tau_S^+=2V_{0S}/\Delta\theta_S$, $\tau_S^-=4V_{0S}/\Delta\theta_S$, where $\tau_S^{\pm}$ is the positive and negative torque on stator potential $V_{1,2}$, respectively, and the extended potential to the left of $P\to I$ has the same slope as $\tau_S^+$.  $V_R$: a symmetric form is used with $V_{0R}=10k_BT$. The transition rate is $k_b=10^5 s^{-1}$ for $\theta_S\ge 2\pi/5$ and $0$ otherwise.}
\label{fig:2}
\end{figure}

In our simulations, we set the stator load $\xi_S$ to be a small constant and change the rotor load $\xi_R$ to obtain the torque and speed for different values of $\xi_R$. The motor dynamics depends on the relative strength between the MotA-FliG coupling interaction potential $V_R$ and the torque-generating potential $V_S$. In Fig.~2, we present the results for a case in which these two potentials are comparable. As shown in Fig.~2A, the torque-speed curve shows a downward concave shape consistent with experiments. In addition to the torque-speed relationship, our model allows us to investigate the rotational dynamics of both the rotor and the stator. In Fig.~2B, we plot the time trajectories of the stator (dashed lines) and rotor (solid lines) angles for different loads. At a low load, the two trajectories are close to each other, whereas at a high load, the trajectory of the rotor is almost flat, even though the stator still rotates at a finite speed. As shown in Fig.~2C, as load increases, the rotor's speed quickly decreases to zero, while the stator's speed approaches a constant speed (reduced but still finite). Our results indicate that the rotations of the stator and the rotor are in-sync at low load: when the stator rotates an angle of $\Delta\theta$, the rotor rotates approximately $k\Delta\theta$, i.e., there is no slippage between the two rotating rings. However, slippage between the two rings increases with the load. Near stall, the rotor speed vanishes while the stator still rotates at a finite speed. The existence of the stator-rotor slippage depends on the strength of the stator-rotor (MotA-FliG) coupling ($V_R$) relative to that of the MotA-MotB coupling ($V_S$). When we used a $V_R$ that is much stronger than $V_S$, slippage disappears. Here, we choose to show a case in which $V_R$ and $V_S$ are comparable to highlight the possibility of slippage between the two rotating protein nano-rings.  


To investigate the relative motion (rotation) of the stator and rotor in details, we plot the trajectories of $\theta_R(t)$ vs. $\theta_S(t)$ for different loads in Fig.~2D, where the dashed black line indicates the tight coupling (no slippage) limit when $\theta_R=k\theta_S$ (dotted black line). At low load (the green line), the trajectory is almost a straight line parallel to the tight coupling limit, while at high load (the red line), the trajectory wiggles closer to the horizontal direction. We can categorize the relative stator-rotor motion by coarse-graining the trajectories in the $\theta_S-\theta_R$ space. We found three types of relative motion, which are shown in the zoom-in plots in Fig.~2D. In type I movement, the rotor moves with the stator without slippage; in type II movement, the rotor-stator system follows a biased random walk in the $(\theta_S,\theta_R)$ space with net motion lower for both the stator and the rotor; in type III movement, the stator rotates at a finite speed but the rotor has almost no net motion, which indicates a high degree of slippage between the rotor and the stator. (Type III movement can be considered as the extreme case of type II movement.) From the results of our simulation, we find that as load increases, the motor spend more time in the type II\&III movements, i.e., slipping. This observation is further confirmed by the distributions of the scaled angle difference between the rotor and stator, i.e., $\varphi=\theta_R-k\theta_S$, as shown in Fig.~2E. At low load, the distribution is concentrated around $\varphi=0$, indicating the absence of slippage. As load increases, the distribution of $\varphi$ becomes more populated in the region with $\varphi<0$, which shows that the stator leaves the rotor behind more often, i.e., the motor spends more time slipping.   




\section{Discussion}
In this article, we propose a new model for BFM based on a general modeling framework and the recently discovered molecular structure of the stator complex. The most important update in our model is that the stator is now modeled as an active rotating nano-machine that consumes PMF and drives the rotation of the rotor. The model can generate the characteristic torque-speed curve observed in experiments. More importantly, our model can be used to investigate and predict the dynamics of both the rotor and the stator, which can be tested in experiments. 

It is generally believed that proton flux and motor rotation are ``tightly coupled" in the BFM, which implies that there is no proton flux at stall and the  efficiency of the BFM is close to $1$ near stall [\cite{HBerg2003}]. Here, we find that if the MotA-FliG interaction is comparable to (or weaker than) the MotA-MotB interaction, there is slippage between the rotor and the stator. The probability of slippage increases as the load increases, which leads to a motor efficiency that is less than $1$ near stall. On the other hand, we also find that slippage can be suppressed by increasing the rotor-stator interaction strength. Thus, our study poses a simple but important question. Does the BFM slip? Ultimately, this question needs to be answered experimentally by measuring the rotation of the stator (MotA-ring) at different loads, especially at high load where the probability of slippage is expected to be the highest if it exists.  

Our model can be extended to cases with multiple stators. 
At low load, the stators and rotor rotate in sync without slippage, so the rotor's maximum speed is determined by the maximum speed of the individual stator units independent of the number of stators. However, at high load (e.g., near stall), the effective duty ratio of the stator, can be far below $1$ due to the higher probability of slippage and/or the relatively long time spent at the energy potential minima waiting for the conformational change [\cite{Meacci2009}]. Note that the effective duty ratio during slipping is zero although the stator is not disengaged with the rotor. Assuming no direct coordination among stators, there should be little overlap among duty cycles of each stator. Thus, the maximum torque at stall is simply the sum of torques generated by all the stators. The dependence of maximum speed and maximum torque on the number of stators have been reported before [\cite{blair1988,yuan2008}], and our model may provide a natural explanation for these experimental observations. 

The rotation of the stator units is also essential for understanding switching between CW and CCW rotation of the rotor. As pointed out in [\cite{santiveri2020}], the MotA ring always rotates clockwise around the MotB dimer, and the switch between the CW and CCW rotation of the rotor ring is determined by conformational changes of the C-ring.  In the CCW mode, the C-ring adopts a compact conformation, and the stator units contact FliG subunits on the outside of the C-ring; thus, the C-ring rotates CCW. Upon binding of CheY-P to FliM in the C-ring, the C-ring expands, and the stator units are now in contact with FliG subunits on the inside of the C-ring, so the C-ring rotates clockwise. In this manner, the unidirectional rotation of stator drives the bidirectional rotation of the motor. In our modeling framework, the coupling between rotor and stator is modeled by $V_R(\theta_R,\theta_S)$. We hypothesize that the CW and CCW rotation modes correspond to different forms of the interaction potential $V_R$, which can lead to the different torque-speed curves for the BFM in CCW and CW rotation observed experimentally [\cite{Yuan2010}]. 

The BFM, like all other biological motors, converts chemical energy (e.g. from ATP hydrolysis or ion translocations) into mechanical work. A fundamental question is whether there are thermodynamic bounds to the power generation and energy efficiency for these highly non-equilibrium molecular engines [\cite{Parmeggiani1999,Parrondo2002,Astumian2010}]. A related question is what microscopic design principles allow a molecular motor to approach these bounds under realistic constraints. In our model, the motor design space is spanned by three functions: $V_S(\theta_S)$, $V_R(\varphi)$ and $k_b(\theta_S)$, which determine the motor dynamics. For example, as we showed in this paper, depending on the relative strength of $V_R$ and $V_S$, there could be slippage between the stator and the rotor, which strongly affects the efficiency of the motor. Each of these functions contains information about the molecular details of the motor, which, in principle, can be estimated from or constrained by the structural data. A systematical exploration of the motor design space will lead to a better understanding of the fundamental limits and possible design principles for optimal motor performance.

\section*{Conflict of Interest Statement}
Yuhai Tu is employed by IBM T. J. Watson Research Center. The remaining authors declare that the research was conducted in the absence of any commercial or financial relationships that could be construed as a potential conflict of interest.

\section*{Author Contributions}

YT originated the study.
YS and TL performed all simulations. YS and YT prepared the manuscript. All authors read and approved the final manuscript.

\section*{Funding}
The work by YT is supported by a NIH grant (R35GM131734). 

\section*{Acknowledgments} YT would like to express deepest thanks to Howard Berg for introducing the bacterial flagellar motor - ``a biotechnological marvel" [\cite{HBerg2003}] to him more than a decade ago and for Howard's constant enlightenment and encouragement on the topic till his untimely passing at the end of 2021. We dedicate this work to the memory of Howard - it was Howard and members of the Berg lab who first told us about the new stator structure [\cite{santiveri2020,deme2020}], which launched this study. YT would also like to thank Nicholas Taylor for an insightful discussion on details of the stator structure and proton translocation process.  

\section*{Data Availability Statement}
No experimental data was used in this study. Inquiries about the simulation results can be directed to the corresponding author.

\section*{Supplementary Material}
There is no Supplementary Material for this article.


\bibliographystyle{naturemag} 
\bibliography{motor_ref}





\end{document}